\documentclass[aps,prl,reprint,superscriptaddress,showpacs,floatfix]{revtex4-1}

\usepackage[draft]{changes}
\usepackage[normalem]{ulem}

\usepackage{graphicx}
\usepackage{amssymb} 
\usepackage{dcolumn}
\usepackage{bm}
\usepackage[mathlines]{lineno}
\usepackage[colorlinks,linkcolor=blue,anchorcolor=blue,citecolor=blue,urlcolor=blue]{hyperref}
\usepackage{multirow}
\usepackage{color}
\usepackage{amsmath}

\makeatletter
\renewcommand*{\@fnsymbol}[1]{\ensuremath{\ifcase#1\or \dagger\or *\or \ddagger\or
\mathsection\or \mathparagraph\or \|\or **\or \dagger\dagger \or
\ddagger\ddagger \else\@ctrerr\fi}} \makeatother

\usepackage{color}

\usepackage[draft]{changes}
\usepackage[normalem]{ulem}

\usepackage{color}%

\begin{document}
\title{
An interpretable formula for lattice thermal conductivity of crystals
}
%

%

\author{Xiaoying Wang}
\affiliation{State Key Laboratory for Mechanical Behavior of Materials, School of Materials Science and Engineering, Xi'an Jiaotong University, Xi'an 710049, China}

\author{Guoyu Shu}
\affiliation{State Key Laboratory for Mechanical Behavior of Materials, School of Materials Science and Engineering, Xi'an Jiaotong University, Xi'an 710049, China}
\author{Guimei Zhu}
\affiliation{School of Microelectronics, Southern University of Science and Technology, 
Shenzhen, 518055, PR  China}

\author{Jian-Sheng Wang}
\affiliation{Department of Physics, National University of Singapore, Singapore 117551, Republic of Singapore}

\author{Jun Sun}
\affiliation{State Key Laboratory for Mechanical Behavior of Materials, School of Materials Science and Engineering, Xi'an Jiaotong University, Xi'an 710049, China}

\author{Xiangdong Ding}
\affiliation{State Key Laboratory for Mechanical Behavior of Materials, School of Materials Science and Engineering, Xi'an Jiaotong University, Xi'an 710049, China}

\author{Baowen Li}
\email[E-mail: ]{libw@sustech.edu.cn}
\affiliation{Department of Materials Science and Engineering, Department of Physics, Southern University of Science and Technology,
Shenzhen 518055, People's Republic of China}

\author{Zhibin Gao}
\email[E-mail: ]{zhibin.gao@xjtu.edu.cn}
\affiliation{State Key Laboratory for Mechanical Behavior of Materials, School of Materials Science and Engineering, Xi'an Jiaotong University, Xi'an 710049, China}

\date{\today}
\begin{abstract}
{Lattice thermal conductivity ($\kappa_L$) is a crucial physical property of crystals with applications in thermal management, 
such as heat dissipation, insulation, and thermoelectric energy conversion. However, accurately and rapidly determining
$\kappa_L$ poses a considerable challenge. In this study, we introduce an formula that achieves high precision (mean relative error=8.97\%) 
and provides fast predictions, taking less than one minute, for $\kappa_L$ across a wide range of inorganic binary and ternary materials. 
Our interpretable, dimensionally aligned and physical grounded formula forecasts $\kappa_L$ values for 4,601 binary and 6,995 ternary materials 
in the Materials Project database. Notably, we predict undiscovered high $\kappa_L$ values for AlBN$_2$ ($\kappa_L$=101 W m$^{-1}$ K$^{-1}$) 
and the undetected low $\kappa_L$ Cs$_2$Se ($\kappa_L$=0.98 W m$^{-1}$ K$^{-1}$) at room temperature. This method for determining
$\kappa_L$ streamlines the traditionally time-consuming process associated with complex phonon physics. It provides insights into microscopic heat transport and facilitates the design and screening of materials with targeted and extreme $\kappa_L$
values through the application of phonon engineering. Our findings offer opportunities for controlling and optimizing macroscopic transport 
properties of materials by engineering their bulk modulus, shear modulus, and Grüneisen parameter.}
\end{abstract}



\maketitle



%

\section{I. Introduction}
Materials with ultrahigh or low thermal conductivity ($\kappa_L$) are desirable for many 
technological applications. 
In the realm of heat dissipation and on-chip cooling, which are
the main challenges for the future development of microelectronics driven by Moore's 
law~\cite{bell2008cooling,li2012colloquium,garimella2008thermal,van2020co}.
Also GaN-based power devices is calling for high $\kappa_L$~\cite{liang2021fabrication,mion2006accurate}. 
Another aspect, thermal insulation and thermoelectric 
technology which enables the direct conversion between heat and electricity requires low $\kappa_L$. 
The conversion
efficiency is determined by the dimensionless figure of merit (\textit{ZT}), which is defined as
$ZT$ = ${S^2 \sigma}$/${(\kappa_e + \kappa_L)}$ (where $\sigma$, $S$, $T$ and $\kappa_e$
represent the electrical conductivity, Seebeck coefficient, absolute temperature and electronic thermal conductivity, respectively). Since the conflicting electronic transport 
properties~\cite{snyder2008complex,he2017advances},
pushing low $\kappa_L$ to the limit of best thermal 
insulators~\cite{gibson2021low} is a favorable strategy to gain 
high \textit{ZT} materials~\cite{zhao2014ultralow}.

So far, there are many traditional
approaches to obtain $\kappa_L$
of inorganic semiconductor materials. %
On the theoretical aspect, one can use  (i) phonon Boltzmann transport 
equation (BTE) with \textit{ab initio} study~\cite{lindsay2010optimized}, 
(ii) molecular dynamics simulations with Green-Kubo or Fourier's 
law~\cite{bao2018review}, (iii) nonequilibrium Green’s 
function~\cite{wang2006nonequilibrium}, and (iv) scattering-matrix~\cite{chen2005effect}. 
For the experiment, there are (i') 
microfabricated suspended thermal bridge~\cite{kim2001thermal,liu2014profiling}, 
(ii') micro-Raman spectroscopy~\cite{balandin2008superior}, (iii') 3$\omega$ 
method~\cite{cahill1987thermal}, and (iv') time-domain 
thermoreflectance (TDTR)~\cite{hu2015spectral}. 

Nevertheless, the theoretical calculation is usually limited 
by insufficient accuracy, time-consuming, and the appropriate 
interatomic force constants (IFCs). The experimental one is always highly 
dependent on leading-edge instruments and the quality of the 
sample~\cite{balandin2011thermal,wei2016intrinsic}. 
To overcome above inherent disadvantages, empirical model and other newly-developed formula based on machine learning (ML) provide new insights~\cite{slack1979thermal}.


It is reported that the strategy of formula based on empirical and  newly-developed technique have made great progress. Empirical methods raised several analytical models for lattice thermal conductivity including the widely-known Slack model~\cite{slack1979thermal} and Debye–Callaway Model~\cite{PhysRev.113.1046}. Furthermore, machine learning (ML) methods have been supplied to obtain the formula for the prediction of thermal conductivity in recently several studies~\cite{chen2019machine,wan2019materials,luo2023predicting}.


However, a difficult-to-obtain directly parameter in the formula, such as Debye temperature $\theta$, hinders the ability to screen and predict thermal conductivity rapidly.
 Furthermore, Machine learning formulas tend to have complex forms and large amount of parameters, which pose a common 
drawback is that most of these models consider physical dimensional 
alignment. Existing a complex formula and an ambiguous physical 
interpretation~\cite{loftis2020lattice,liu2020high}.
Moreover, more and more investigation testified that calculating the thermal conductivity requires consideration of higher order scattering, we put high-order scattering factor $\delta$ into considering.



In this work, we first collect all available $\kappa_L$ materials from the AFLOW database~\cite{curtarolo2012aflow}.
Then we detect some key physical quantities as a function of $\kappa_L$ aided by Pearson correlation analysis (Fig.~\ref{fig2}). We further relate these parameters to elastic mechanics, phonon-phonon scattering, relaxation time approximation, and domain knowledge, simplifying the Slack formula with a more precise and compact form written as Eq. (\ref{finalfinal}). Comparison of the formula's calculation results with  Slack formula and density functional theory (DFT) is conducted in order to confirm its accuracy. For the application of formula, considering the ``small data'' characteristics of Grüneisen parameter $\gamma$, we solve it through crystal graph convolutional neural networks (CGCNN) as illustrated in Fig.~\ref{fig5}(c). Finally we apply our formula for predicting 
thermoelectric materials. The overview of entire process is provided in Fig.~\ref{fig1}.

Our proposal has a universal and simple empirical formula with strong generalization ability and clear physical meaning. Our formula, Eq. (\ref{finalfinal}), coupled with the machine learning method, is not only accurate 
but also able to predict $\kappa_L$ rapidly. We finally apply our
method to 4,601 binary and 6,995 ternary inorganic compounds in the Materials 
Project database, which can accelerate the materials discovery 
process with targeted $\kappa_L$ and reduce the design and screening 
costs in thermal management, on-chip cooling, and thermoelectrics.

\section{A surrogate formula for lattice thermal conductivity}

\begin{figure*}
\centering
\includegraphics[width=1.6\columnwidth]{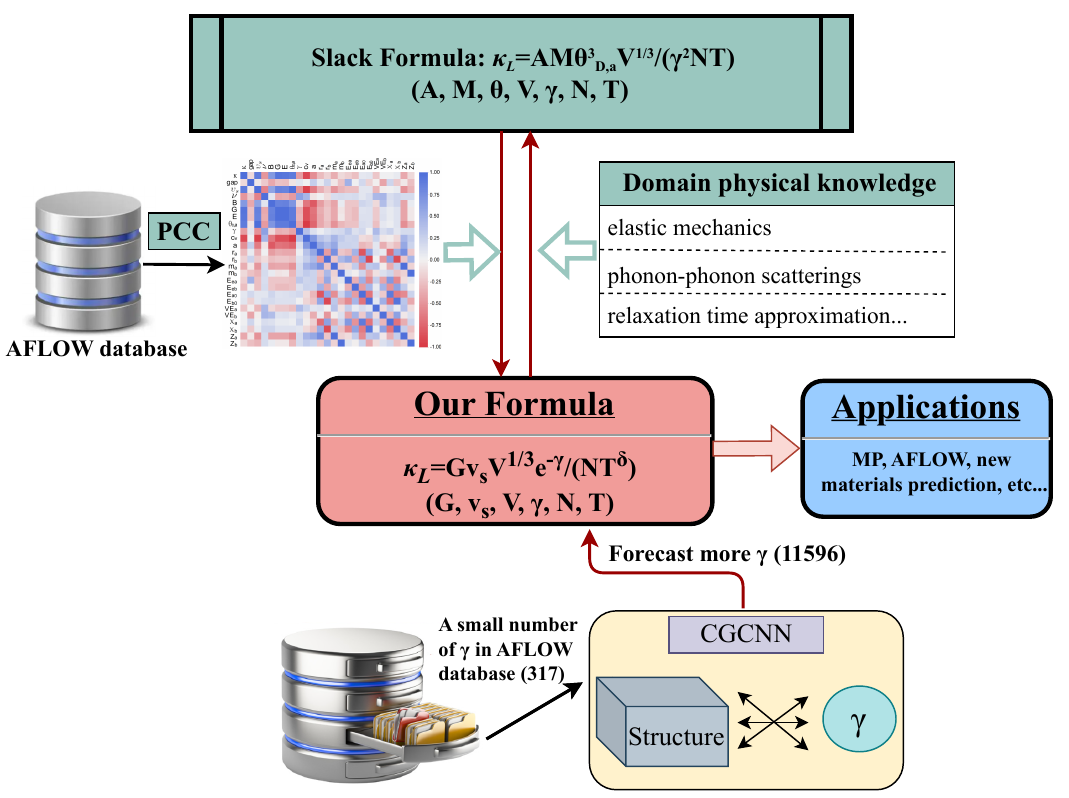}
\caption{
The schematic framework of the proposed approach. In the Slack model, $M$ is the atomic mass of the atom and $V$ is the volume of 
the primitive cell. $\theta_{D,a}$ is the Debye temperature of 
acoustic phonons. $\gamma$, $N$, and $T$ are Grüneisen parameters, 
the number of atoms in the primitive cell, and absolute temperature, 
respectively. The added parameters G and $\upsilon_s$ in our formula represent the shear modulus and the phonon velocity of sound, respectively. In the process of formula derivation, $\upsilon_s$ can be replaced by $\rho$, $B$ (Bulk modulus) and $G$ according to Eq. (\ref{velocity}). 
We screened a total 317 number of $\gamma$ data from the ALFOW database and predicted a total of 11596 $\gamma$ data through crystal graph
convolutional neural network (CGCNN).
We used Pearson correlation coefficients (PCC) to analyze correlated 
parameters and via domain knowledge, derive an interpretable formula. We finally apply our method to all binary and ternary inorganic compounds in the Materials Project and AFLOW
database, which can accelerate the materials discovery process with targeted $\kappa_L$.}
\label{fig1}
\end{figure*}

In this section, we derive our key finding of a more concise and accurate formula for predicting thermal conductivity, Eq. (\ref{finalfinal}), from analysis of Pearson correlation coefficients  (PCC) and some physics arguments.

\begin{figure}
\centering
\includegraphics[width=1.0\columnwidth]{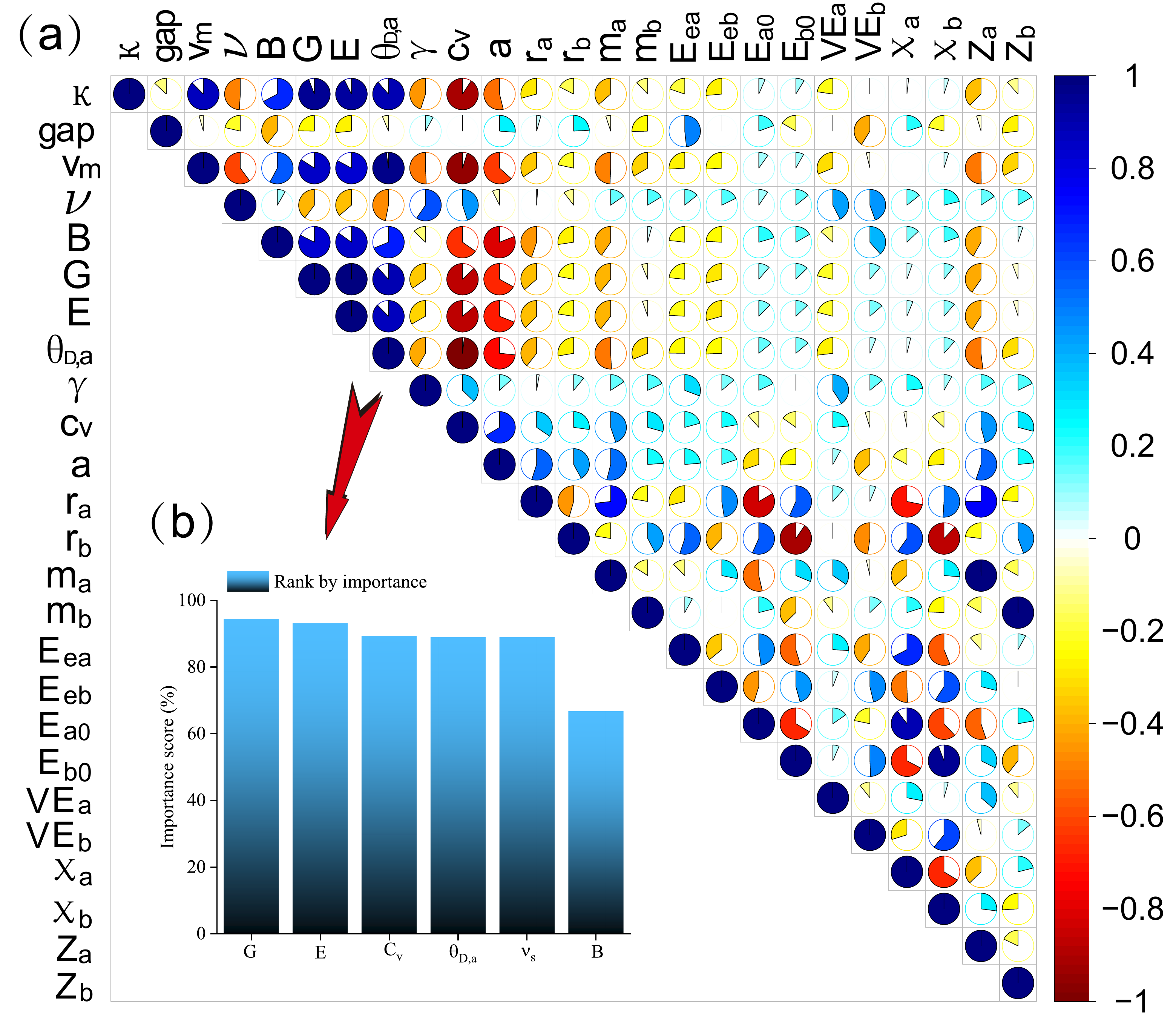}
\caption{(a) Pearson correlation coefficient (PCC) between lattice thermal conductivity $\kappa_L$,
electronic band gap, speed of sound $\upsilon_s$, Poisson’s ratio $\nu$, bulk modulus $B$,
shear modulus $G$, Young’s modulus $E$, Debye temperature $\theta$$_{D,a}$, Grüneisen 
parameter $\gamma$, heat capacity $C_V$, lattice constant $a$, their atomic radius
$r_a$, $r_b$, their atomic mass $m_a$, $m_b$, their electron affinity $E_{ea}$,
$E_{eb}$, their ground state energy per atoms $E_{a0}$, $E_{b0}$, their number of 
valence electrons $VE_a$, $VE_b$, their electronegativities $\chi$$_a$, $\chi$$_b$,
their atomic number of atoms $Z_a$, $Z_b$ for inorganic compounds crystals. 
The blue section is a positive 
correlation, while the red one is a negative correlation. 
(b) The importance ranking of features via the PCC method is shown at the left bottom. 
} %
\label{fig2}
\end{figure}

Due to a large number of parameters affecting $\kappa_L$, we initially 
use the PCC which is defined in Supplemental Material, to select critical 
parameters and proper descriptors~\cite{PhysRevX.6.041061}. Fig.~\ref{fig2}(a)
displays the correlogram for whole inorganic compounds crystals from AFLOW 
database~\cite{curtarolo2012aflow} between the following variables: 
lattice thermal conductivity $\kappa_L$, electronic band gap, speed 
of sound $\upsilon_s$, Poisson’s ratio $\nu$, bulk modulus $B$, shear 
modulus $G$, Young’s modulus $E$, Debye temperature $\theta$$_{D,a}$, %
Grüneisen parameter $\gamma$, heat capacity $C_V$, lattice 
constant $a$, their atomic radius $r_a$, $r_b$, their atomic mass 
$m_a$, $m_b$, their electron affinity $E_{ea}$, $E_{eb}$, their 
ground state energy per atoms $E_{a0}$, $E_{b0}$, their number of 
valence electrons $V E_a$, $V E_b$, their electronegativities 
$\chi$$_a$, $\chi$$_b$, and their atomic number $Z_a$, $Z_b$. %

It is interesting to note that $G$, $E$, $C_V$, $\theta$$_{D,a}$, 
$\upsilon_s$, and $B$ are largely correlated with thermal conductivity 
$\kappa_L$, shown in Fig.~\ref{fig2}(b). The correlation value 
between these parameters and $\kappa_L$ is almost more than 70\% 
by PCC analysis. $G$, $E$, $B$ and $\upsilon_s$ are the physical 
quantities of elastic mechanics, while $C_V$ and $\theta$$_{D,a}$
are parameters of thermodynamics. For the Peierls-Boltzmann 
transport equation, the lattice thermal conductivity $\kappa_L$
can be calculated as
\begin{equation} %
\label{kappa}
\kappa_{\alpha\beta} = \frac{1} {V} \sum_{\lambda} C_\lambda
                       \upsilon_{\lambda \alpha}
                       \upsilon_{\lambda \beta} \tau_{\lambda},  \\
\end{equation}
where \textit{V} is the crystal volume, $\lambda$ denotes a phonon mode 
with a different wave vector and branch index. $\alpha$ and $\beta$ 
denotes the Cartesian directions. \textit{C$_\lambda$} is the specific 
heat per mode, $\upsilon$$_{\lambda \alpha}$ and $\tau$$_{\lambda}$ 
are the velocity component along $\alpha$ direction and the relaxation
time of the phonon mode $\lambda$. Therefore, $\kappa_L$ is a coupling
effect between harmonic IFCs and higher-order ($\geq$3) 
IFCs~\cite{gao2018unusually}. Note that PCC is a linear correlation 
analysis but $\kappa_L$, especially phonon relaxation time $\tau$, 
originates from the higher-order phonon-phonon 
scattering~\cite{wang2023role}. Therefore, we should clarify the 
relationship between these physical quantities. Besides, the 
Grüneisen parameter $\gamma$ has been shown to significantly 
affect the value of thermal expansion and 
$\kappa_L$~\cite{Morelli2006}, even if its PCC value is less 
than 50\% shown in Fig.~\ref{fig1}(a). Therefore, based on the 
PCC analysis and domain knowledge, $G$, $E$, $B$, 
$\upsilon_s$ (harmonic effect), $C_V$, $\theta$$_{D,a}$, and 
$\gamma$ (anharmonic effect) are selected as the most critical 
feature descriptors for $\kappa_L$.

Various early estimates 
of $\kappa_L$ of a solid at temperatures not too far removed from the Debye 
temperature has been discussed by Slack~\cite{slack1979thermal}. This 
estimate takes the form,
%
\begin{equation}
\kappa_L=A\frac{M\theta_{D,a}^3V^{\frac{1}{3}}}{\gamma^2 NT},
\label{early model}
\end{equation}
where $M$ is the atomic mass of the atom and $V$ is the volume of 
the primitive cell. $\theta_{D,a}$ is the Debye temperature of 
acoustic phonons. $\gamma$, $N$, and $T$ are Grüneisen parameters, 
the number of atoms in the primitive cell, and absolute temperature, 
respectively. There are two kinds of expression of the parameter $A$, one
is $A(\gamma)=\frac{2.43\times 10^{-8}}{(1-\frac{0.514}{\gamma}+\frac{0.228}{\gamma^2})\gamma^2}$
given by Julian~\cite{julian1965theory}, and other one is 
$A=3.04\times10^{-8}$ obtained by Slack~\cite{slack1979thermal}. 
Since $\gamma$ in most materials 
can not reach 2 and is a variable quantity rather than a constant 
number, we use a proportional expression that is dimensionally 
aligned, %

\begin{equation}
\kappa_L\propto\frac{M \theta_{D,a}^3 V^{\frac{1}{3}}}{NT}.
\label{propto2}
\end{equation}

Combing the traditional definition for Debye temperature $\theta_{D}=\frac{\hbar \upsilon_s}{k_B}(\frac{3}{4 \pi}\frac{N}{V} )^{\frac{1}{3}}$  and derived directly from the phonon density of states by integrating only over the acoustic modes $\theta_{D,a}= \theta_D/N^{\frac{1}{3}}$~\cite{toher2017combining},
then $\theta_{D,a}$ can be expressed as Eq. (\ref{thetaDa})
\begin{equation}
\theta_{D,a}=\frac{h}{k_B}\left(\frac{3}{4\pi} \cdot \frac{1}{V}\right)^{\frac{1}{3}}\upsilon_s,
\label{thetaDa}
\end{equation}

Basically, $\upsilon_s$ is the speed of sound in the elastic mechanics and 
can be calculated by the transverse velocity $\upsilon_t$ and the 
longitudinal velocity $\upsilon_l$~\cite{jia2017lattice}, %
\begin{align}
    \upsilon_s&=\left\{\frac{1}{3}\left[\frac{1}{\upsilon_l^3}+\frac{2}{\upsilon_t^3}\right]\right\}^{-\frac{1}{3}}\nonumber, \\
    \upsilon_l&=\sqrt{\frac{B+\frac{4}{3}G}{\rho}} \label{velocity},\\
    \upsilon_t&=\sqrt{\frac{G}{\rho}}\nonumber,
\end{align}
\label{vs}

Inserting Eq. (\ref{thetaDa}) into Eq. (\ref{propto2}), the lattice thermal conductivity 
can be transformed into a new form,
\begin{equation}
\kappa_L \propto \frac{3h^3}{4 \pi k_B^3 } \cdot \frac{M}{V} \cdot \frac{ V^{\frac{1}{3}} \upsilon_s^3} {NT},
\label{proto33}
\end{equation}

One can remove the constant terms and attain,
\begin{equation}
\kappa_L\propto\left(\rho \upsilon_s^2 \right)\left(\frac{\upsilon_s V^{\frac{1}{3}}}{NT} \right),
\label{proto3}
\end{equation}
where $\rho$ is the density of a material. 
$\kappa_L$
is proportional to the hardness as expressed in  the preceding study, a single crystal diamond has a 4100 W m$^{-1}$ K$^{-1}$ at 
104 K~\cite{wei1993thermal}. Besides, carbon nanotube~\cite{fujii2005measuring} and silicon 
carbide~\cite{morelli2002estimation} are also good cases. Consequently, 
it is assumed that $\kappa_L$ is proportional to the (Vickers) hardness 
of materials. $\kappa_L\propto H_V$, it is 
intuitive to assume that they deliver similar information about $\kappa_L$. 
Therefore, one can use hardness $H_V$ to replace $\rho \upsilon_s^2$,
\begin{equation}
\kappa_L \propto H_V \cdot \left(\frac{\upsilon_s V^{\frac{1}{3}}}{NT}\right),
\label{slack4}
\end{equation}

A renowned work~\cite{teter1998computational} 
found that $H_V$ is rigorously proportional to shear modulus $G$, rather than 
Young's modulus $E$. Thus, neglecting the contribution of 
plasticity~\cite{CHEN20111275}, hardness $H_V$ can be precisely estimated 
by~\cite{teter1998computational}, 
\begin{equation}\label{teter}
    H_V=0.151\cdot G,
\end{equation}
%
%
and $\kappa_L$ can be expressed as,
\begin{equation}
    \kappa_L\propto\frac{G \upsilon_s V^{\frac{1}{3}}}{NT} \cdot B(\gamma),
    \label{slack5}
\end{equation}
where $B(\gamma)$ is a dimensionless term associated with Grüneisen 
parameter $\gamma$. 
A question arises 
naturally: what is the relationship between B($\gamma$) and phonon 
relaxation time? According to the work of Klemens, the inverse 
relaxation time of phonons for U-type process can be written 
as~\cite{klemens1958thermal},
\begin{equation}
\tau^{-1}_U=B_U\omega T^3 \cdot {\rm exp} (-\frac{\theta_D}{\alpha T}),
\label{klemens}
\end{equation}
where phonon relaxation time $\tau$ and Debye temperature $e^{\theta_D}$ 
is a positive correlation. Besides, $\theta_D= \omega_{max}/{k_B}$ and 
$\gamma=- d{\rm ln}\omega/{d{\rm ln}V}$, combing Eq.(\ref{klemens}), one 
can conclude that $\tau$ is negatively proportional to $\gamma$. Moreover,
they are likely to have an exponential relationship. Based on the above
theoretical derivation and domain knowledge~\cite{klemens1958thermal}, we 
propose that $B(\gamma)=e^{-\gamma}$ and the final thermal conductivity is,
\begin{equation}
\kappa_L=\frac{G \upsilon_s V^{\frac{1}{3}}}{NT} \cdot e^{-\gamma},
\label{final}
\end{equation}

Note that Eq. (\ref{final}), in consistent with the Slack model of Eq. (\ref{early model}),
 only consider the contribution from acoustic phonon branches, which is good for common materials, in which group velocities of optical branches are small and the acoustic branch dominates the  heat transport. 
 Therefore, for other materials wherein the optical branches play a critical role, the formula needs to be revised, for example, BaO  whose $\kappa_L$ is estimated 
of 14.15 W m$^{-1}$ K$^{-1}$~\cite{surplice1963thermal}, while the 
experimental value is 2.3 W m$^{-1}$ K$^{-1}$~\cite{surplice1963thermal}.  
We revise the rate of $\kappa_L$ decline by,
\begin{equation}
\kappa_L=\frac{G \upsilon_s V^{\frac{1}{3}}}{N T^\delta} \cdot e^{-\gamma}.
\label{finalfinal}
\end{equation}
where $\delta$ is somewhere between 1 and 2. The precise theory of the power
law is quite complex, having to do with competition between scattering 
processes produced by cubic and quartic anharmonic 
terms~\cite{herring1954role,wang2023role}. In the following, we only consider 
three-phonon scattering with $\delta = 1$. Therefore, combining the PCC analysis and physical domain knowledge,
we derive an interpretable and dimensionally aligned formula for lattice thermal conductivity.

Our formula, Eq. (\ref{finalfinal}) has several advantages compared to the Slack model, Eq. (\ref{early model}):
(i) The Debye temperature ($\theta_D$) is replaced by the Shear modulus ($G$) and sound velocity ($\upsilon_s$).
These two parameters are more conveniently obtained and searchable from databases such as the Materials Project and AFLOW. 
(ii) The exponential form of our Grüneisen parameter ($\gamma$) incorporates Kelemens's well-known relaxation time approximation
for U-type phonon-phonon scattering processes in Eq. (\ref{klemens}), compared with the experimental fit of $\gamma^2$ in the Slack model.
(iii) Our formula can incorporate fourth-order phonon scattering using the $T^{\delta}$ form, where $\delta$ is somewhere between 1 and 2.
%

\begin{figure*}
\centering
\includegraphics[width=1.5\columnwidth]{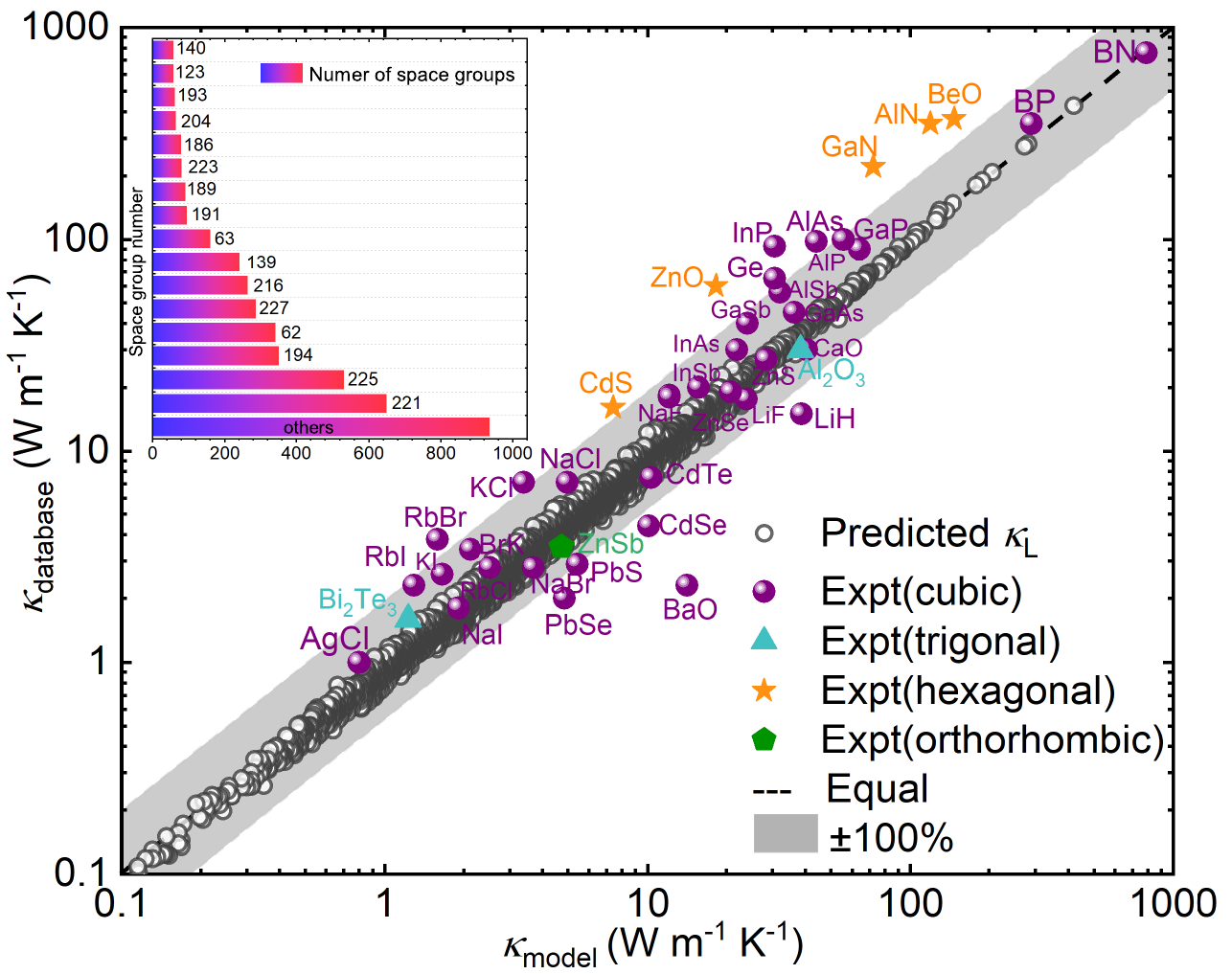}
\caption{Comparison of thermal conductivity $\kappa_L$ at 300 K between our formula prediction
and ALFOW database~\cite{curtarolo2012aflow} with gray hollow points. The 
purple, cyan, orange, and green colors stand for experimental 
measurement (AgCl~\cite{beasley1994thermal}, Al$_2$O$_3$~\cite{slack1962thermal}, 
BaO~\cite{surplice1963thermal}, ZnSb~\cite{bjerg2014modeling}, Bi$_2$Te$_3$,
CdSe~\cite{toberer2011phonon}, AlAs, AlN, AlP, AlSb, BeO, BN, BP, CaO, CdS,
CdTe, GaAs, GaN, GaP, GaSb, Ge, InAs, InP, InSb, KBr, KCl, KI, LiH, LiF, NaBr,
NaCl, NaF, NaI, PbS, PbSe, PbBr, RbCl, RbI, ZnO, ZnS, ZnSe, and ZnTe~\cite{Morelli2006}) 
with cubic, trigonal, hexagonal, and orthorhombic crystal systems, respectively.
The dashed black line is the function of $y=x$. The inset in the upper left 
corner represents the distribution of the number of space groups (the horizontal 
axis) for all materials. The vertical axis represents the index of the space group.
}
\label{fig3}
\end{figure*}

\section{Formula validation with the Slack model}

In order to verify the accuracy of our formula with Slack model, we collect the whole 
ALFOW compounds which was calculated by the Slack model. Downloadable databases of $\kappa_L$ include 4363
single, binary, and ternary inorganic materials. 
The result is shown in Fig.~\ref{fig3}. The 
horizontal axis represents predicted $\kappa_L$ based on the 
Eq. (\ref{final}) and the vertical axis plots the $\kappa_L$ 
originating in the AFLOW database~\cite{curtarolo2012aflow}. The black 
data points are essentially concentrated on the range of the $y = x$ 
function, indicating a high degree of proximity for our formula. We also
list 42 familiar materials from four crystal systems to verify 
Eq. (\ref{final}), coupled with the experimental data shown in 
the Supplemental Material. The data distribution is shown in the inset
marked with the number of space groups.

Meanwhile, our formula can also be used for different crystal systems, as 
shown in Fig.~\ref{fig3} with different symbols and colors. The error analysis displays that the majority of materials are surrounded 
within 100\% in gray color. We also plot 
the corresponding experimental measurement. It should be pointed out that the 
data are distributed in various space groups as shown in inset, illustrating that our formula
has an effective prediction for various crystal structures and element 
independence. It is noteworthy from Fig.~\ref{fig3} that our model 
achieves a high accuracy of our proposed framework and provides a novel 
methodology for designing unusual materials with targeted properties, 
such as ultrahigh and ultralow $\kappa_L$. Besides, the whole computational
cost, compared to the \textit{ab initio} study, is almost negligible.

 Besides, we not only can predict the AFLOW 
database but also apply it to other new materials. Here, we adopt the 
precursor Materials Project to make a prediction. We estimate $\gamma$
and $\kappa_L$ based on our model for 4,601 binary and 6,995 ternary 
inorganic materials from the Materials Project. All results are shown
in the Supplemental Material MP-binary and MP-ternary. Part of their $\kappa_L$ was not published to the best of our knowledge. Therefore, 
our model can be a good predictor to compare, rank, and design the unknown 
materials. %

\begin{table*}
\caption{Four examples of materials from the Materials Project database. 
$\rho$ (g$\cdot$cm$^{-3}$), $V$ (\AA$^3$), $G$ (GPa), $B$ (GPa), and $N$ 
represent density, volume, Shear modulus, Bulk modulus, and number of atoms in the primitive cell, respectively. $\kappa_{DFT}$ and $\kappa_{model}$ (W m$^{-1}$ K$^{-1}$) 
stand for lattice thermal conductivity from the ShengBTE code and 
our proposed model from Eq. (\ref{finalfinal}). Except for SiC, 
AlBN$_2$, Cs$_2$Se, and AgCl are novel materials that have never 
been reported about their $\kappa_L$.\label{thermal}}
\centering
\begin{tabular}{{c}{c}{c}{c}{c}{c}{c}{c}{c}}
\hline
Materials  & ID-number   & $\rho$  & $V$   & $G$   & $B$   & $N$  & $\kappa_{DFT}$  & $\kappa_{model}$   \\
\colrule
SiC     & mp-8062  & 3.17 & 21.00   & 187 & 211 & 2 & 420.65       & 369.53              \\ 
AlBN$_2$      & mp-1008557  & 3.33 & 32.82  & 177  & 258  & 4  & 101.54       & 146.81    \\ 
Cs$_2$Se      & mp-1011695 & 3.73 & 153.89  & 5   & 12  & 3  & 0.98        & 0.96          \\
AgCl      & mp-22922 & 5.58 & 44.43  & 7   & 43  & 2     & 0.36          & 0.39             \\ 
\botrule
\end{tabular}
\end{table*}

\begin{figure*}
\centering
\includegraphics[width=1.0\textwidth]{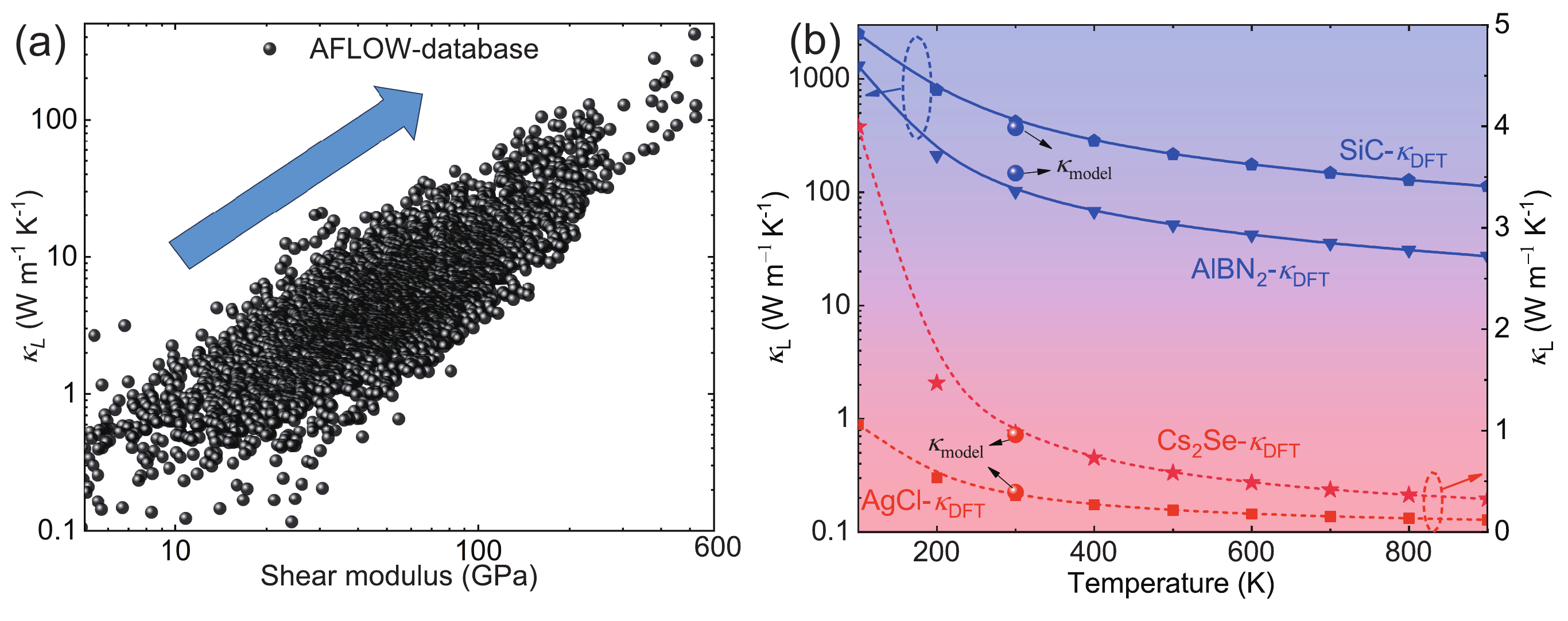}
\caption{%
(a) The relationship of Shear modulus $G$ and $\kappa_L$ originated from the AFLOW database.
(b) Example of ultrahigh and ultralow thermal conductivity $\kappa_L$ materials 
between our formula predictions (solid symbols) and \textit{ab initio} 
calculations (solid and dashed lines). The Grüneisen parameter $\gamma$
of 300 K was substituted into our formula to obtain $\kappa_L$ represented
as bullets.}
\label{fig4}
\end{figure*}

\section{Accuracy comparison between our formula and DFT study} 

Combining our empirical model of high-throughput computation for predicting $\kappa_L$, we demonstrate the empirical formula as a function of the Shear modulus predicted by our model (Fig.~\ref{fig4}(a)). Mapping relation exists in $G$ with $\kappa_L$ visibly. It is proved that $G$ plays a significant role in thermal conductivity and our model is truthful.

Afterwards, we verify the accuracy of our theoretical model by randomly selecting two kinds of high $\kappa_L$ materials (blue dotted lines)
and two kinds of low $\kappa_L$ materials (red dotted lines) compared with the \textit{ab initio} density functional theory (DFT) calculation, shown in Fig.~\ref{fig4}(b).
The physical parameters 
output from the selected materials are shown in Table~\ref{thermal}.

In Fig.~\ref{fig4}(b) and Table~\ref{thermal}, 
the Grüneisen parameter $\gamma$ of SiC is 0.66 at 300 K. Coupled with $G$ (187 GPa) 
and $B$ (211 GPa), one can obtain $\kappa_{model}$ of 369.53 W m$^{-1}$ K$^{-1}$ based 
on our model. Well match with solving the phonon BTE by DFT calculation, SiC has a value of 
420.65 W m$^{-1}$ K$^{-1}$ at room temperature. Similarly, for also high $\kappa$ 
material, AlBN$_2$ has a $\gamma$ of 0.94. By substituting $\gamma$ into 
Eq. (\ref{finalfinal}), we can get a $\kappa_{model}$ is 146.81 W m$^{-1}$ K$^{-1}$ 
at 300 K, while the \textit{ab initio} study is 101.54 W m$^{-1}$ K$^{-1}$. 
The result shows that our model has a small error compared with the DFT 
calculation for high thermal conductivity materials. On the other hand, for the low
thermal conductivity materials, $\gamma$ of Cs$_2$Se and AgCl are 1.41 and 2.61 
and the predicted $\kappa_{model}$ are 0.96 W m$^{-1}$ K$^{-1}$ and 
0.39 W m$^{-1}$ K$^{-1}$ at 300 K, respectively. 
For the phonon BTE calculation, 
$\kappa_{DFT}$ are 0.98 W m$^{-1}$ K$^{-1}$ and 0.36 W m$^{-1}$ K$^{-1}$ of both materials.

In this sense, our model is more accurate for low thermal conductivity materials than
high $\kappa_L$ materials. But all prediction results and DFT results are of 
the same order of magnitude, which implies our model can be used for screening 
and designing targeted $\kappa_L$ materials with tunable density, volume, Bulk 
modulus, and number of atoms in the primitive cell by phonon engineering~\cite{qian2021phonon}. %

%

\begin{figure*}
\centering
\includegraphics[width=1.0\textwidth]{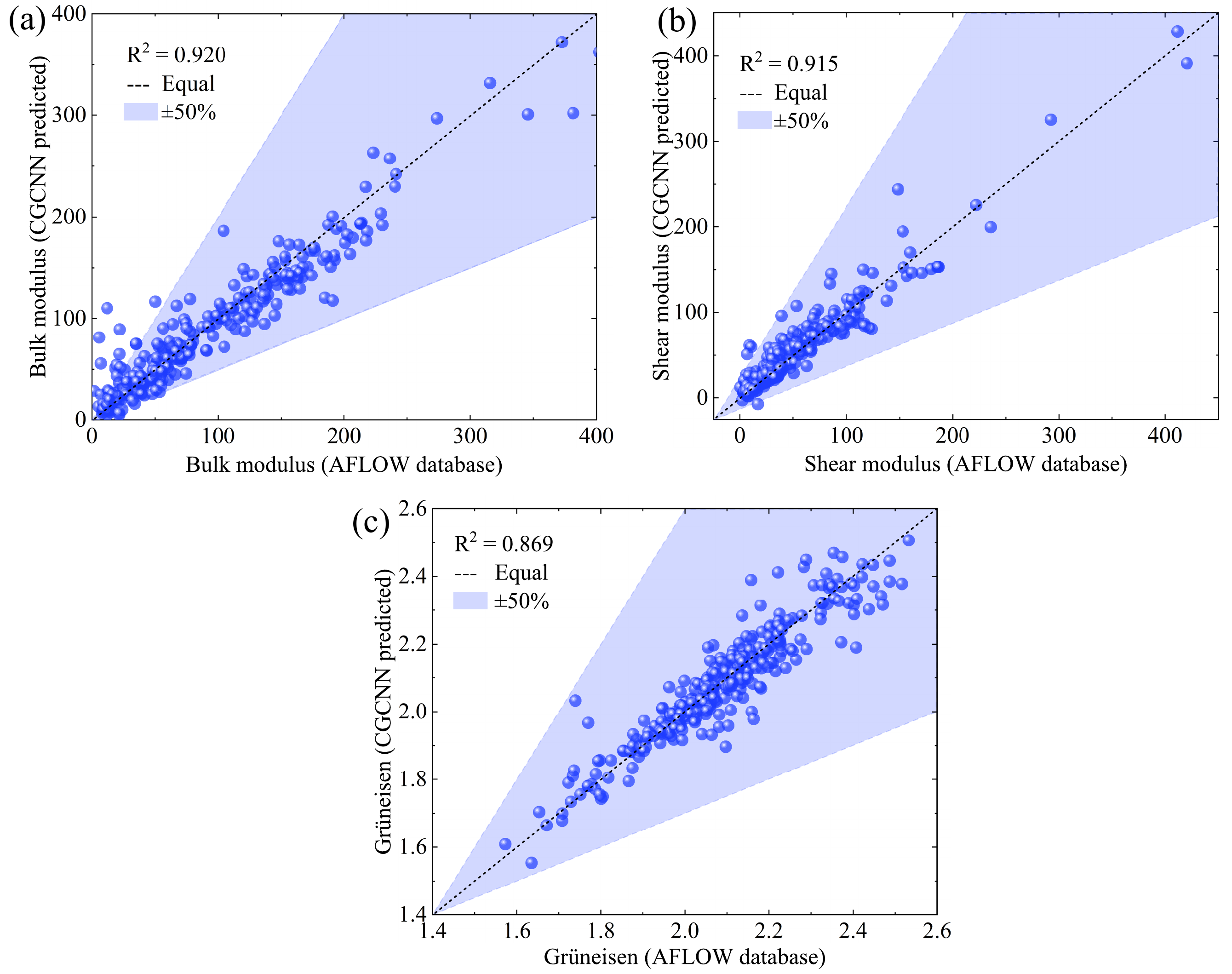}
\caption{%
(a) Based on the CGCNN framework, the correlation of Bulk modulus (B) between our predicted values and the AFLOW database. (b) The correlation of Shear modulus (G) between our predicted values and the AFLOW database. (c) Comparison of $\gamma$ between our trained-CGCNN prediction
and the benchmark of the AFLOW database~\cite{curtarolo2012aflow}.}
\label{fig5}
\end{figure*}

\section{Machine learning aids formula parameter $\gamma$}

In brief, in order to acquire  the thermal conductivity $\kappa_L$, 
except for the crystal structure acquiring volume of the primitive cell and the number of atoms, there 
are only three parameters that need to be known: $B$ (so as to obtain $\upsilon_s$ according to Eq. (\ref{velocity})), $G$, and $\gamma$.
$\gamma$ can normally be estimated by the thermal expansion data. The 
calculation of $\gamma$ needs high computational cost~\cite{gao2018unusually}, using the trained CGCNN
method is a fast and efficient method of obtaining large quantities of $\gamma$ ~\cite{xie2018crystal}. 
$B$ and $G$ are elastic results but $\gamma$ are inelastic response. Therefore, linear 
response parameters $B$ and $G$ are easier to obtain than $\gamma$ which is nonlinear.

\begin{figure*}
\centering
\includegraphics[width=1.0\textwidth]{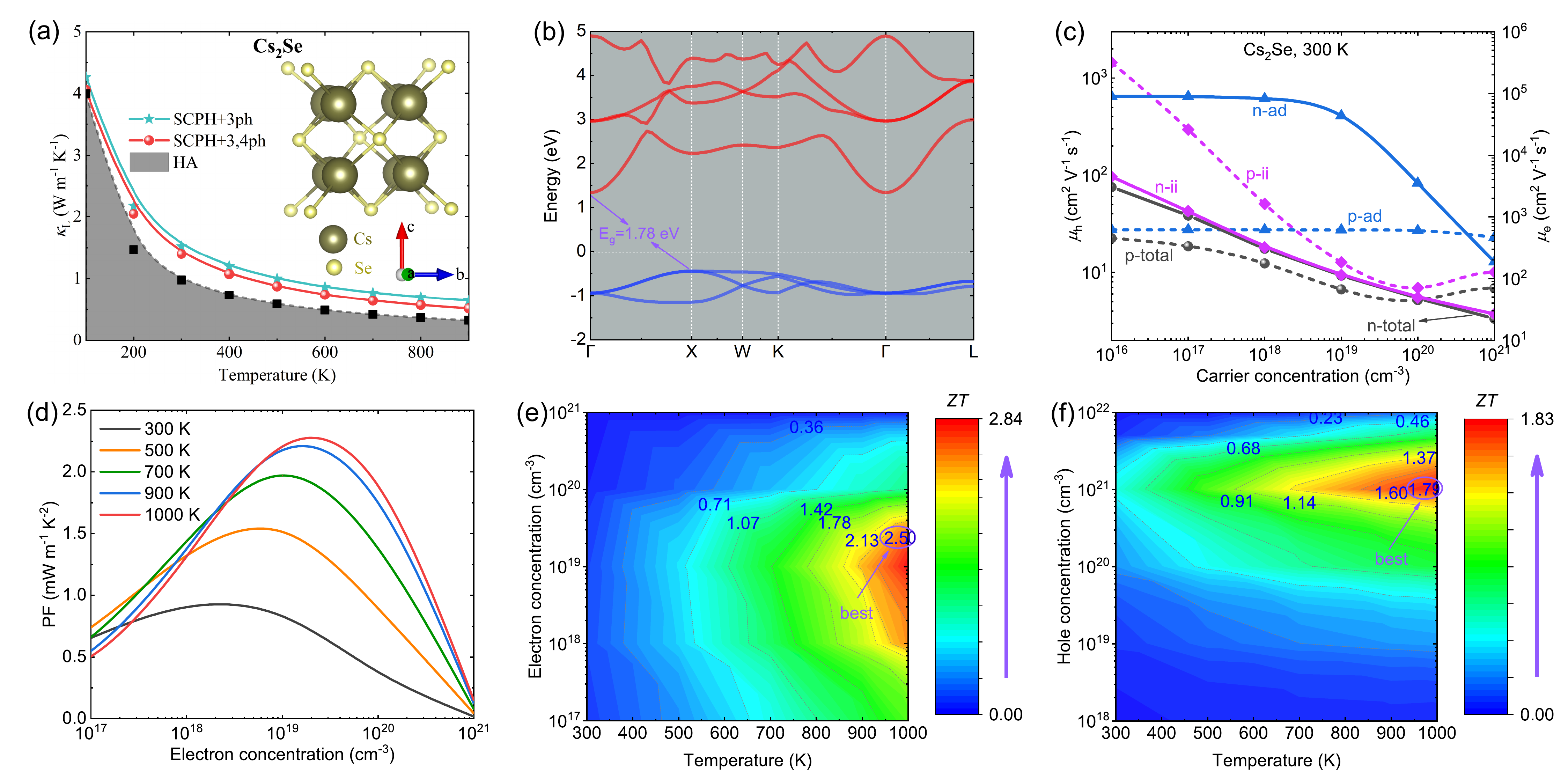}
\caption{%
The calculated transport performance for predicted Cs$_2$Se novel materials. %
(a) The computed lattice thermal conductivity of harmonic approximation (HA), third-order anharmonicity phonon renormalization (SCPH+3ph), and quartic anharmonicity phonon renormalization (SCPH+3,4ph) of Cs$_2$Se as a function of temperature.
(b) Electronic band structure with conduction bands (red) and valence 
bands (blue), respectively. The electronic band gap is 1.78 eV of Cs$_2$Se. %
(c) Hole p-type mobility with dotted lines (left ordinate) and electronic 
n-type mobility with solid lines (right ordinate) as a function of carrier 
concentration. Scattering mechanisms not only consider the acoustic 
deformation (ad) potential scattering (blue) but also include ionized 
impurity (ii) scattering (purple). The total mobility is marked in black
solid (n-total) and dashed (p-total) lines as a function of carrier 
concentration. (d) Power factor $PF=S$$^2$$\sigma$ for n-type Cs$_2$Se 
electronic transport at 300 K, 500 K, 700 K, 900 K, and 1000 K, respectively. %
Contour plot of thermoelectric figure of merit \textit{ZT} for (e) n-type and (f)
p-type Cs$_2$Se as a function of temperature and carrier concentration. The lattice thermal conductivity 
$\kappa_L$ is calculated by self-consistent (SCPH) theory and the four-phonon 
scattering mechanism.}
\label{fig6}
\end{figure*}

In order to provide more support for conducting thermal conductivity calculations based on $B$ and $G$ data.
We use the CGCNN method to predicted the 
$B$ and $G$, shown in the Fig.~\ref{fig5}(a) and Fig.~\ref{fig5}(b). 
Furthermore, we obtained all existing $\gamma$ values of binary and ternary compounds
from the AFLOW database~\cite{curtarolo2012aflow} and used them as a training 
process to build a nonlinear mapping network between the crystal structure
and phonon anharmonicity $\gamma$ as shown in Fig.~\ref{fig5}(c).

$R^2$ (root mean square error) 
evaluates the accuracy of $\gamma$ between the predicted values and 
benchmark from the AFLOW database.
$R^2$ for $B$, $G$, and $\gamma$ are 0.920, 
0.915, and 0.869, respectively.
The model performance of $\gamma$ is comparatively trustworthy and can be 
used for subsequent $\kappa_L$ prediction based on Eq. (\ref{finalfinal}).
We supply predicted $\gamma$ into our formula associating with existing renowned database Material Project, totally 4,601 binary and
6,995 ternary novel materials. As displayed in Supplemental Material (MP-binary and MP-ternary files).

\section{Prediction of a new thermoelectric material}

On account of low $\kappa_L$ is a favorable factor to gain high $ZT$ materials and best thermal insulators. In order to expand the application of our formula, we select one unreported low $\kappa_L$ material, Cs$_2$Se, as an example, to explore the  potential thermoelectric performance.

The phonon and electronic transport properties of Cs$_2$Se are shown in Fig.~\ref{fig6}.
Based on quasi-harmonic approximation, we notice that our predicted $\kappa_L$ 
agreed with the DFT calculation. However, recent works show that four-phonon
scattering and quartic anharmonicity play a critical role in heat transport
~\cite{tadano2015self,xia2020high,wang2023role}. 
We label this advanced method as abbreviated SCPH+3ph, SCPH+3,4ph.
We adopt the rigorous self-consistent phonon (SCPH) approximation which comprises 
the first-order contribution to the phonon self-energy and temperature-dependent 
phonons~\cite{tadano2015self,xia2020high}. 3ph represents three-phonon is taken into account. 3,4ph means both three-phonon and four-phonon scattering are considered. 
 More details of Cs$_2$Se $\kappa_L$ can be 
found in Fig.~\ref{fig6}(a). 


Due to the temperature renormalization and stiffening of phonon dispersion, $\kappa_L$  (SCPH+3ph) is larger than $\kappa_L$  (SCPH+3,4ph). Moreover, the quartic anharmonicity induces extra phonon-phonon scattering, leading to a smaller $\kappa_L$  (SCPH+3,4ph) compared with $\kappa_L$  (SCPH+3ph).


All energy bands are shifted to the Fermi level at 0 eV shown in Fig.~\ref{fig6}(b). 
It has an indirect band gap of 1.78 eV. 
Renowned SnSe is a layered 
crystal with a band gap of 0.86 eV~\cite{zhao2014ultralow}. 
Recent work shows that operating  at a high temperature is a 
priority for high band-gap materials~\cite{xiao2020seeking}. 
The best thermoelectric performance at the optimal working temperature is 
restricted by the band gap ($E_g$) owing to $E_g = 2e S_{max} T$, where $e$
is the unit charge, $S_{max}$ is the maximum Seebeck coefficient, and $T$
is the temperature that corresponds to $S_{max}$. In this sense, wide-band
gap semiconductors may work over a wide temperature range and the \textit{ZT} 
values are not saturated at high temperatures. %

Fig.~\ref{fig6}(c) plots the holes (p-type in blue) and electrons (n-type in purple) 
mobility of Cs$_2$Se at 300 K. On the basis of electronic BTE, the electronic relaxation 
time ($\tau$) must be given to obtain the mobility $\mu$ and electric conductivity 
$\sigma$. On the one hand, some researchers will approximately adopt a constant 
relaxation time of 10-12 fs~\cite{he2016ultralow}.
Therefore, the total 
mobility is expressed as 1/$\tau$$_{total}$ = 1/$\tau$$_{ad}$ + 1/$\tau$$_{ii}$. Both
type of mobility decreases as a function of carrier concentration due to the enhanced 
electron-phonon scattering. $n$-type total mobility is much higher than $p$-type at the 
same carrier concentration. %

Because the electric conductivity increases with the carrier concentration, the Seebeck 
coefficient decreases with the concentration, so there is an optimal doping for the power 
factor ($PF=S$$^2$$\sigma$). The n-type power factor increase as a function of temperature,
while the p-type $PF$ is reduces with the temperature increasing as shown in the Supplemental 
Material Fig. S9. The largest $n$-type 
power factor is 2.3 mW m$^{-1}$ K$^{-2}$ for Cs$_2$Se at 1000 K, shown in Fig.~\ref{fig6}(d). 
The largest p-type power factor is 3.4 mW m$^{-1}$ K$^{-2}$ at 300 K. 
From Fig.~\ref{fig6}(b), the valence band maximum (VBM)
is flatter than conduction band minimum (CBM), resulting in a ``pudding-mold'' which consists
of a dispersive portion and a flat portion, can in general be favorable for the coexistence 
of large Seebeck coefficient and small resistivity~\cite{kuroki2007pudding}. Therefore, $p$-type
Cs$_2$Se has a larger power factor compared with $n$-type. %

With all of the above results of phonon transport electron transport properties, 
\textit{ZT} can be determined, as shown in contour plot Fig.~\ref{fig6}(e) and 
Fig.~\ref{fig6}(f). Cs$_2$Se displays the largest $ZT$ value of 2.50 (1.79) 
for $n$-type ($p$-type) at 1000 K. Such a high \textit{ZT} value is comparable 
to the 2.6 $\pm$ 0.3 at 923 K in SnSe~\cite{zhao2014ultralow}
 and other Copper ion liquid-like 
thermoelectrics~\cite{liu2012copper}. Our calculation suggests that Cs$_2$Se 
is a promising material for thermoelectric applications, especially for 
operating at a high temperature range and also provides a guidance for unveiling 
new thermoelectric materials based on low $\kappa_L$. %

\section{CONCLUSIONS}
In summary,  building upon the Slack model, we derived an enhanced formula for 
quickly and conveniently calculating lattice thermal conductivity, incorporating 
relaxation time approximation and fourth-order phonon scattering. Our findings 
include: (i) a strong correlation between lattice thermal conductivity, Shear 
modulus, speed of sound, and Grüneisen parameter;  (ii) the development of an 
interpretable, fast, and accurate theoretical formula for lattice thermal 
conductivity, Eq. (\ref{finalfinal}), by integrating fundamental physical 
principles with machine learning; (iii) the potential use of the theoretical 
formula, Eq. (\ref{finalfinal}), in discovering new materials for heat dissipation, 
thermoelectrics, and refrigeration, characterized by ultralow or ultrahigh thermal 
conductivity

\section{ACKNOWLEDGMENTS}
The authors gratefully acknowledge discussions with Lucas Lindsay, Natalio Mingo, Wu Li.
We acknowledge the support from the National Natural Science Foundation of China 
(No.12104356 and No.52250191). 
Z.G. acknowledges the support of China Postdoctoral Science Foundation (No.2022M712552), %
the Opening Project of Shanghai Key Laboratory of Special Artificial Microstructure Materials 
and Technology (No.Ammt2022B-1), and the Fundamental Research Funds for the Central 
Universities. 
The work is supported by the Key Research and Development Program of the Ministry of 
Science and Technology under Grant No.2023YFB4604100.
We also acknowledge the support by HPC Platform, Xi’an Jiaotong University.



 \bibliography{References} 
 \end{document}